\documentclass[journal,transmag]{IEEEtran}

\usepackage{amsmath}

\usepackage{algorithmic}

\usepackage{array}

\usepackage{dblfloatfix}
\usepackage{url}
\usepackage{booktabs}
\usepackage{tikz}
\usepackage{makecell}
\usepackage{threeparttable}
\usepackage{longtable}
\usepackage{multirow}
\usepackage{tabularx}
\usepackage{tabulary}
\usepackage{tablefootnote}
\usepackage{amssymb}
\usepackage{epstopdf}
\usepackage{graphicx}
\usepackage{subfig}
\usepackage{caption}
\usepackage{color}
\usepackage{mathtools}
\usepackage{hyperref}
\usepackage{bm}
\usepackage{dsfont}
\usepackage{enumitem}
\begin{document}

\title{Disentangled Multimodal Representation Learning for Recommendation }

\author{Fan~Liu,~\IEEEmembership{Member,~IEEE}, Huilin~Chen, Zhiyong~Cheng,
         Anan~Liu, Liqiang~Nie,~\IEEEmembership{Senior Member,~IEEE} and \\
        Mohan~Kankanhalli,~\IEEEmembership{Fellow,~IEEE}
	\IEEEcompsocitemizethanks{\IEEEcompsocthanksitem F. Liu and M. Kankanhalli are with School of Computing, National University of Singapore, Singapore. Email: liufancs@gmail.com, mohan@comp.nus.edu.sg
	\IEEEcompsocthanksitem H. Chen is with the School of Computer Science and Engineering, Tianjin University of Technology, China. Email:ClownClumsy@outlook.com
	\IEEEcompsocthanksitem Z. Cheng is with the Shandong Artificial Intelligence Institute, Qilu University of Technology (Shandong Academy of Sciences), China.
	Email: jason.zy.cheng@gmail.com
	\IEEEcompsocthanksitem A. Liu is with the School of Electrical and Information Engineering, Tianjin University, China. Email:anan0422@gmail.com
	\IEEEcompsocthanksitem 
	L. Nie is a professor with Harbin Institute of Technology (Shenzhen), China. Email: nieliqiang@gmail.com
	}
	\thanks{Fan Liu is the corresponding author.}
    }

\markboth{IEEE TRANSACTIONS ON MULTIMEDIA}%
{Shell \MakeLowercase{\textit{et al.}}: Bare Demo of IEEEtran.cls for Computer Society Journals}

\IEEEtitleabstractindextext{
\begin{abstract}
Many multimodal recommender systems have been proposed to exploit the rich side information associated with users or items (e.g., user reviews and item images) for learning better user and item representations to improve the recommendation performance. Studies from psychology show that users have individual differences in the utilization of various modalities for organizing information. Therefore, for a certain factor of an item (such as \emph{appearance} or \emph{quality}), the features of different modalities are of varying importance to a user. However, existing methods ignore the fact that different modalities contribute differently towards a user’s preference on various factors of an item. In light of this, in this paper, we propose a novel \emph{Disentangled Multimodal Representation Learning} (DMRL) recommendation model, which can capture users’ attention to different modalities on each factor in user preference modeling. In particular, we employ a disentangled representation technique to ensure the features of different factors in each modality are independent of each other. A multimodal attention mechanism is then designed to capture users' modality preference for each factor. Based on the estimated weights obtained by the attention mechanism, we make recommendations by combining the preference scores of a user’s preferences to each factor of the target item over different modalities. Extensive evaluation on five real-world datasets demonstrate the superiority of our method compared with existing methods.
\end{abstract}

\begin{IEEEkeywords}
Multimodal, Modality Preference, Disentangled Representation, Recommendation 
\end{IEEEkeywords}}

\maketitle
\IEEEdisplaynontitleabstractindextext
\IEEEpeerreviewmaketitle

\section{Introduction}

\IEEEPARstart{R}{ecommender} systems have been widely applied in modern web services, such as e-commerce platforms, news websites and social media sites. Recommendations not only assist users in quickly finding the desired content from a massive amount of information, but also help develop the  providers (e.g., Amazon~\footnote{https://www.amazon.com}, eBay~\footnote{https://www.eBay.com}, TikTok~\footnote{https://www.tiktok.com\label{tiktok}}) services. Among various recommendation techniques, Collaborative Filtering (CF) based methods~\cite{Koren2009MF,he2017neural} have been very successful, attributed to its simple idea of learning the user and item representations by exploiting the interaction data (e.g., \emph{click} or \emph{purchase}). 
However, a problem of only using the interaction data is that the performance will dramatically drop when the interactions between users and items are sparse.
An effective way to overcome this problem is to leverage the  side information associated with users or items, which provide rich information about user preference and item characteristics~\cite{mcauley2015image,cheng2016wide,chen2018attention}. 


So far, there have been many approaches proposed to exploit various types of side information, such as attributes~\cite{cheng2016wide,chen2018attention}, reviews~\cite{catherine2017transnets,mcauley2013hidden,tan2016rating,cheng2018aspect,chin2018anr} and images~\cite{mcauley2015image,wang2017your}. Most existing methods often explore the capability of a single modality, like reviews~\cite{cheng2018aspect,chin2018anr} or images~\cite{mcauley2015image,wang2017your}, on enhancing the recommendation performance. In the multimedia domain, it has been well-recognized that features from different modalities are complementary to each other and the use of multi-modal features can  help improve the performance in general~\cite{guo2018multi,Hehe2020Tomm,Fan_Yang_2020}. In recent years, researchers have also exploited the multimodal information for learning more comprehensive user and item representation in recommendation systems~\cite{zhang2016collaborative,zhang2017joint,hsieh2017cml,liu2018MAML,wei2019mm,Wei2019GRCN,DingMWC22}. 
Based on how multimodal information is utilised in user and item embedding learning, we broadly classify the existing methods into two types. 
The first type integrates the features learned from different modalities via a linear~\cite{zhang2016collaborative,wei2019mm,Wei2019GRCN} or non-linear method~\cite{zhang2017joint,liu2018MAML} to obtain a joint representation for each user or item. For instance, JRL~\cite{zhang2017joint} concatenates the representations learned from ratings, reviews and images to form the final representation. The other type utilizes a regularization scheme to constrain the relations between representations of different modalities in a latent space~\cite{hsieh2017cml,liu2018MAML}. For example, CML~\cite{hsieh2017cml} uses the multimodal features as a regularization term with a designed loss to facilitate the embedding learning of items and users.

It is well-known that users preferences on items are diverse. That is, the cause for a user to like an item are varied across different items. For example, a user may favor a product due to its \emph{quality} and \emph{brand}, while prefer another one just because of its \emph{aesthetic appearance}. Because multimodal information provides rich features that directly describe different aspects and even evidences of users' opinion on those aspects (e.g., comments), it has also been exploited to model the diverse preferences of users towards various items. To achieve this, existing methods either learn an attention vector for different aspects with respect to the target user-item pair~\cite{zhang2017joint} or update the target item/user vector based on the multimodal features of the target items~\cite{liu2018MAML,cheng2018aspect,chin2018anr}. The success of the aforementioned methods is largely attributed to their effectiveness of exploiting multimodal information to learn better user and item representations. We argue that they have not fully utilized the multimodal information. This is because they do not disentangle the multimodal features to mine the relative contributions of different modality features to users' preference for each factor of an item.

Firstly, the salient factors captured by different modalities are different. For example, user reviews contain more of users' opinions on  factors such as \emph{quality} and \emph{comfort}~\cite{cheng2018aspect,chin2018anr}. In contrast, item images capture more of user preferences concerning visual appearance (like \emph{color} and \emph{type})~\cite{mcauley2015image,he2016vbpr}. Existing multimodal CF models usually exploit the features of multimodal information without distinguishing the feature differences between different modalities~\cite{zhang2016collaborative,zhang2017joint,guo2018multi}. For example, TranSearch~\cite{guo2018multi} uses a deep neural network to fuse multimodal representations using a non-linear function. The previous studies have not considered the differing contributions of multimodal information at the factor-level for users' preference modeling. 

In addition, users have individual differences in perceiving different modalities for organizing information as pointed out in psychology studies~\cite{Jeffrey1978modalpreferences}. Accordingly, the features of different modalities are of different importance to a user's preference on a certain factor of a target item. In other words, users have different preferences for different modalities (or \emph{modality preference}) in each factor of an item.  For example, the \emph{style} information can be directly viewed from the images or obtained from the descriptions in reviews when purchasing \emph{clothes}. For this factor, some users directly view the images to check whether they like them, while some others might be more willing to refer to the opinions in reviews. However, existing multimodal recommendation methods have not considered  the modality preferences of users in the utilization of information when exploiting different modalities, resulting in sub-optimal performance. 

Finally, the entangled representations of different factors in each modality  can hurt the performance~\cite{Higgins2017betaVAELB,Ma2019DisenGCN,wang2020DGCF}. For example, a user purchases a dress because of its \emph{style} while caring less about \emph{comfort}; otherwise, the user would have purchased the other dress considering both \emph{style} and \emph{comfort}. When the representations of different factors in each modality are entangled, it results in information redundancy and performance degradation in preference modeling. We would like to model user preference from different factors and each factor can be represented by the features of different modalities. Disentangled representation has been demonstrated to be more resilient to feature complexity~\cite{Higgins2017betaVAELB,Ma2019DisenGCN}. Recently, researchers have also applied disentangled representation learning techniques to model users' diverse intents~\cite{wang2020DGCF}. Although some progress has been achieved, they only use the user-item interaction data and leave the rich information contained in multimodal data unused. In fact, the multimodal data provides a lot of useful information to help us discover users' intents on items. The problem is how to effectively disentangle and fuse the multimodal features to better serve our recommendation model. 


Motivated by the above considerations, in this paper, we propose a \emph{Disentangled Multimodal Representation Learning} (DMRL for short) model, which models user's preference by considering the different contributions of features from different modalities for each disentangled factor. In our model, a user's preference for an item is predicted by  aggregating her weighted preferences on all modality features of different factors. To enable robust and independent representations for each factor, we employ a disentangled representation technique to ensure that the features of different factors in each modality are independent of each other. Based on the disentangled representations of different factors, we design a multimodal attention mechanism to capture users’ modality preference on different modalities for each factor. Finally, for the preference prediction, given a user and a target item, we first estimate the user’s preference score for each factor in each modality, and then linearly combine the scores of all the factors across different modalities using the estimated weights obtained by the attention mechanism. To this end, our model can profile users’ personal preference based on disentangled factors represented by multimodalities with the consideration of  their personal modality preference. Extensive experiments on five real-world datasets have been carried out to demonstrate the effectiveness of our method with comparison to several strong baselines, including the ones using multimodal information and disentangled learning techniques.

\section{related work}
\label{sec:related_work}
In this section, we briefly review the recent advances in model-based collaborative filtering methods, especially the ones based on multimodal information and disentangled representation learning, which are closely related to our work. 

In CF models, users and items are represented as dense vectors (i.e., embeddings) in the same latent space. Based on the learned vectors, an interaction function is used to predict the preference of a user to an item. Take Matrix Factorization (MF) as an example, the user and item embeddings are learned by minimizing the error of re-constructing the user-item interaction matrix, and the dot product is used as the interaction function for prediction. This simple idea has achieved tremendous success, and many variants of MF have been developed~\cite{Salak2007PMF,rendle2009bpr,he2017neural,wang2021clicks,wang2022causal}. The advent of deep learning has accelerated the development of model-based CF techniques. Due to the powerful capability of deep learning, it has been widely applied to learn better user and item embeddings~\cite{xue2017deep,Cheng2022FLA,Liu2022SDG} or model more complicated interactions between users and items~\cite{he2017neural,Li2019CNR}. For example, NeuMF~\cite{he2017neural} models the complex interactions between users and items using nonlinear neural networks as the interaction function. 
More recently, Graph Convolution Network (GCN) techniques have also been deployed that have set a new standard for recommendation~\cite{berg2019gcmc,wang2019ngcf,Liu2020A2GCN,Liu2021IMP_GCN}. The advantage of GCN-based recommendation models is attributed to its capability of explicitly modeling higher-order proximity between users and items. For example, 
NGCF~\cite{wang2019ngcf} exploits higher-order proximity by propagating embeddings on the user-item interaction graph.

All of the aforementioned methods learn the representations of users and items merely relying on the interaction data. And their performance suffers when the interactions are sparse. Besides the interaction data, the rich side information, such as reviews and images, provide valuable information of user preference and item characteristics, which has been widely used to alleviate the data sparsity problem in recommendation~\cite{mcauley2013hidden,cheng2016wide,tan2016rating,chen2018attention}. 

\subsection{Multimodal Collaborative Filtering}
Various types of side information have been utilized in recommender systems, such as attributes~\cite{cheng2016wide,chen2018attention}, reviews~\cite{mcauley2013hidden,tan2016rating,cheng2018aspect,chin2018anr} and images~\cite{mcauley2015image,wang2017your}. Most existing models exploits different types of information individually, because of the challenges of fusing multimodal information. However, it is well-recognized that information from a different modality can provide complementary information, as demonstrated in~\cite{zhang2016collaborative,zhang2017joint,liu2018MAML,Wei2019GRCN,Min2020TMM,Cai2021TMM}.

More recently, several deep models have been proposed to model user preference by using multimodal features. Zhang et al.~\cite{zhang2016collaborative} proposed a knowledge based method, which extracts the multimodal knowledge, unstructured textual and visual knowledge to jointly learn the latent representations of items within a collaborative filtering framework. In JRL~\cite{zhang2017joint}, Zhang et al. first extracted user and item features from ratings, reviews, and images separately with deep neural networks, and then concatenated those features to form the final user and item representations for recommendation. For capturing fine-grained user preference, Liu et al.~\cite{liu2018MAML} proposed a metric learning based method, which models diverse user preferences by exploiting the item's multimodal feature. In recent years, Graph Convolution Networks (GCNs) have attracted increasing attention in multimedia recommendation due to the powerful capability on representation learning from non-Euclidean structures~\cite{wei2019mm,Wei2019GRCN,Zhang2021MLS}. MMGCN~\cite{wei2019mm} learns the model-specific user preference to the content information via  direct information interchange between user and item in each modality. Later on, Wei et al.~\cite{Wei2019GRCN} proposed a structure-refined GCN model for multimedia recommendation. 

Despite the progress, the factors behind multimodal features are entangled in representations, resulting in sub-optimal performance.
In this paper, in order to generate robust multimodal representations, we employ the disentangled representation technique to encourage the independence of different features.  


\subsection{Disentangled Representation Learning}
Disentangled representation learning has gained considerable attention, in particular in the field of image representation learning~\cite{Higgins2017betaVAELB}. It aims to identify and disentangle the underlying explanatory factors behind the data. Such representations have demonstrated to be more resilient to the complex variants~\cite{Ma2019DisenGCN}. 

In recent few years, modeling user's diverse intent for liking items have attracted increasing attention. Cheng et al.~\cite{cheng2018aspect} applied topic model on reviews to analyze user preferences on different aspects of items, which are used to model a user's diverse intents to different items. Liu et al.~\cite{liu2018MAML} presented a metric-learning based recommendation model, which employs an attentive neural network to estimate user attention on different aspects of the target item by exploiting the item's multimodal features (e.g., review and image). 

In fact, the factors behind features in the aforementioned methods are highly entangled, which results in sub-optimal performance of recommendation. For learning robust and independent representations from user-item interaction data, disentangled representation are increasingly being valued in recommendation~\cite{ma2019learning,wang2020DisenHAN,wang2020DGCF,Li2022DisenGNN}. Ma et al.~\cite{ma2019learning} captured user preferences regarding the different concepts associated with user intentions separately. Wang et al.~\cite{wang2020DisenHAN} proposed a disentangled heterogeneous graph attention network, which learns disentangled user/item representations from different aspects in a heterogeneous information network. For studying the diversity of user intents on adopting the items, Wang et al.~\cite{wang2020DGCF} presented a GCN-based model, which yields disentangled representations by modeling a distribution over intents for each user-item interaction. 

The above methods apply the disentangled learning techniques on interaction data to model users' diverse intents on adopting items without exploiting any side information. We claim that side information, especially the combination of multimodal information (e.g., reviews and images), contain rich evidence about users' preferences on items. In addition, we claim that for a specific factor, the features of different modalities convey different information. Inspired by this consideration, in this paper, we apply the disentangled learning technique on multimodal information to learn a users' preference on each factor of an item. Moreover, we design a multimodal attention network to capture a user's modality preferences on different modalities for each factor.  

\section{Our Model}
\label{sec:methods}
\subsection{Preliminaries}
\begin{figure*}[t]
	\centering
	\includegraphics[width=0.9\linewidth]{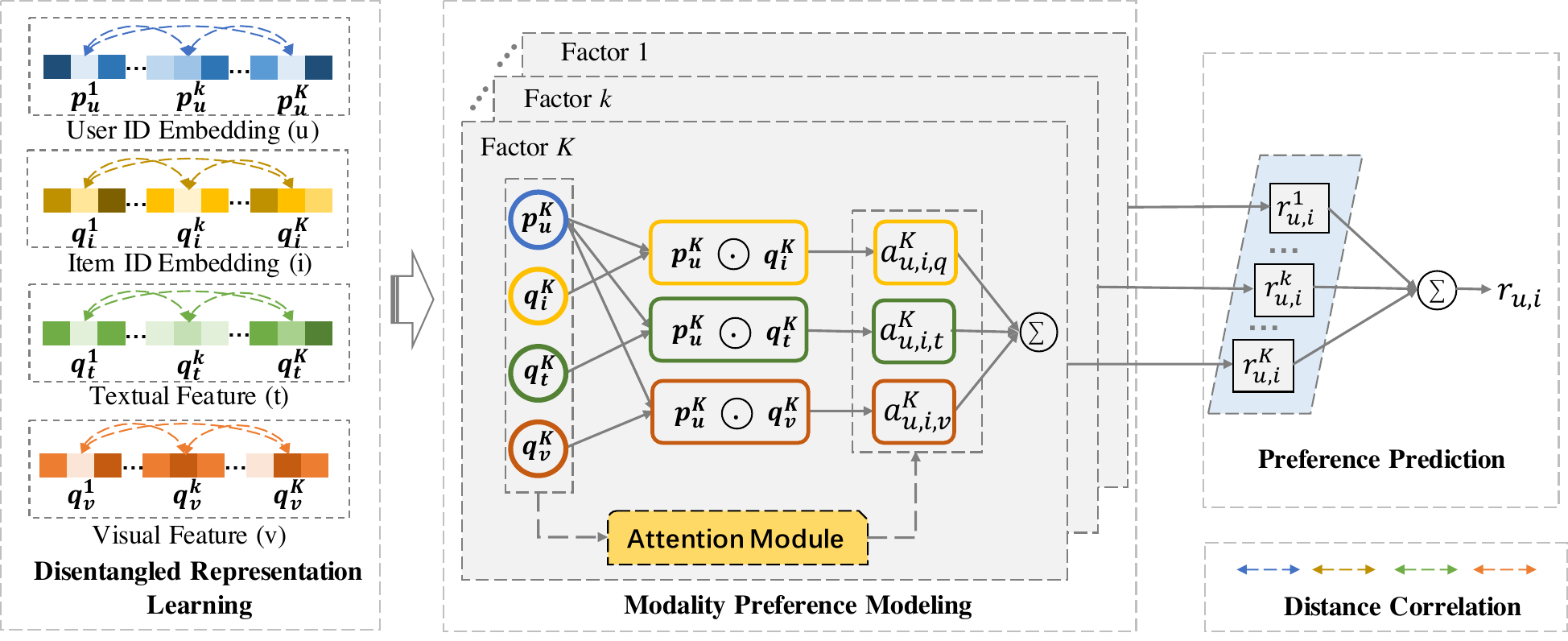}
	\caption{Overview of our DMRL model. Best viewed in color.}
	\vspace{0pt}
	\label{fig:tripartite_graph}
\end{figure*}

Before describing our model, we would like to introduce the problem setting first. Given a user set $\mathcal{U}$ and an item set $\mathcal{I}$, we use $\bm{R}^{N_u \times N_i}$ to denote the user-item interaction matrix, in which a nonzero entry $r_{u,i}\in \bm{R}$ indicates that user $u \in \mathcal{U}$  has interacted with  item ${i} \in \mathcal{I}$ before; otherwise, the entry is zero. Notice that the interactions can be implicit (e.g., click) or explicit (e.g., rating). $N_u$ and $N_i$ are the numbers of users and items, respectively. In our setting, each user and each item is assigned a unique ID, and it is represented by a trait vector. Besides, each item is associated with multi-modal information, e.g., reviews and images. And for each item, its associated review and image are represented as a textual feature vector and a visual feature vector, respectively. It should be noted that item ID, image and review are regarded as three types of modality in our setting.
Our goal is to recommend a user $u\in \mathcal{U}$ with suitable items which the user did not purchase before and will find appealing.

Existing multimodal methods often represent each modality (i.e., ID, review, and image) of each item as a holistic vector. As mentioned earlier, we would like to model users' diverse preference on different factors, and more specifically, their attention to different modalities of the same factor. To achieve this, in our model, the vector of each modality is composed of K chunks, with the assumption that each chunk represents a factor (such as appearance or quality). Without loss of generality, we use the user ID embedding as an example, it is initialized as:
\begin{equation}
	\bm{p_u} = \bm{(p_u^1, p_u^2,\cdots,p_u^K)},
\end{equation}
where $\bm{p_u} \in \mathds{R}^d$ is ID embedding to capture intrinsic characteristics of user $u$; $\bm{p_u^k} \in \mathds{R}^{\frac{d}{k}}$ is $u$'s chunked representation of the $k$-th factor. Analogously, $\bm{q_i} = \bm{(q_i^1, q_i^2,\cdots,q_i^K)}$, $\bm{q_t} = \bm{(q_t^1, q_t^2,\cdots,q_t^K)}$ and $\bm{q_v} = \bm{(q_v^1, q_v^2,\cdots,q_v^K)}$ are defined as the embeddings of item $i$'s ID, textual feature and visual feature, respectively. 

The embeddings of user ID and item ID are randomly initialized in our model, and the initial embeddings of textual and visual features of items are extracted from reviews and images, which are described in the following. 

\textbf{Feature Extraction.} In our implementation, the BERT~\cite{Devlin2019BERTPO} model is utilized to extract the text feature. Specifically, BERT learns continuous distributed vector representations for textual documents
in an unsupervised manner. BERT tries to preserve the semantic information and also considers the sequential information in the word sequence. It takes text documents as inputs and outputs their vector representations in a latent semantic space. The visual feature $\bm{F}_{v}$ is extracted by the ViT ~\cite{Google2021ViT} model that uses the transformer encoder with 32 layers. In this work, we take the output of the transformer encoder, resulting in a 1024-D feature vector as the visual features for each item~\footnote{In this paper, each item is associated with one image.}.

The extracted textual and visual features are then fed into two separate two-layer neural networks, which are used as refinement networks to learn valuable features oriented for recommendation~\footnote{Experiments show that the neural network with two layers could achieve better performance.}. Specifically, the two-layer neural network is:
\begin{equation}
	\bm{q}_{x} = \sigma(\bm{W}_{x}^0\sigma( \bm{W}_{x}^1\bm{F}_{x} + \bm{b}_{x}^1 ) + \bm{b}_x^0),    
\end{equation}
where $\bm{F_x}$ is the input feature vector of the $x$-modality (i.e., visual or textual). $\bm{W}_{x}^0$, $\bm{W}_{x}^1$ and $\bm{b}_{x}^0$, $\bm{b}_{x}^1$ are the corresponding weight matrices and bias vectors for two layers, respectively. $\sigma(\cdot)$ is the activation function and LeakyReLU is employed because of its biologically plausibility and non-saturated property~\cite{wang2019ngcf}. The textual feature $\bm{F_t}$ and visual feature $\bm{F_v}$ are processed by two separate networks to obtain the embeddings of $\bm{q_t}$ and $\bm{q_v}$ respectively for subsequent learning. Noted that our focus in this paper is to study whether the disentangled representation of factors can help us enhance the recommendation performance. The above textual and visual extraction methods can be replaced by any other more advanced methods. In the following, we give the details of our disentangled multimodal representation learning recommendation model.

\vspace{-0pt}
\subsection{Our Model}
As shown in Figure~\ref{fig:tripartite_graph}, our \emph{Disentangled Multimodal Representation Learning} (DMRL) consists of three key components: 1) \textbf{disentangled representation learning}, which employs distance correlation~\cite{Gabor2007M_dis_cov,Gabor2009distance_cov} as a regularizer to encourage the independence of the features of different factors in each modality; 2) \textbf{modality preference modeling}, which captures user modality preference on different modalities for each factor; 3) \textbf{preference prediction}, which predicts the preference of a user to the target item by aggregating the learned disentangled features from different modalities with assigned attention weights. Next, we introduce each component in sequence.

\subsubsection{\textbf{Disentangled Representation Learning}}
We would like to divide the feature representations of each modality into k chunks. And each chunk represents a latent factor which impacts user's preference on the item~\footnote{Note that we use the semantic factors, such as \emph{appearance} and \emph{quality}, as examples for ease of understanding of our model. Latent factors are implicit and unexplainable.}.  For simplicity, the features from each modality are evenly divided into $k$ continuous chunks. As a result, the features of different factors will be entangled. This will cause information redundancy and reduce the efficacy of  preference modeling based on those factors. To avoid this problem, we need to first disentangle the features of different factors in each modality. To achieve this goal, we apply distance correlation~\cite{Gabor2009distance_cov,Gabor2007M_dis_cov} to characterize independence of any two-paired embeddings of different factors. Its coefficient is zero if and only if these vectors are independent. For a feature vector $y$, we formulate it as:
\begin{equation}
	L_{y} = \sum_{k=1}^{K}\sum_{k^{'}=k+1}^{K}dCor\left ( \bm{y}^{k},\bm{y^{k^{'}}} \right ),
\end{equation}
where $dCor(\cdot)$ is the function of distance correlation and it is defined as:
\begin{equation}
	dCor\left ( \bm{y^k},\bm{y^{k^{'}}} \right ) = \frac{dCov\left ( \bm{y^{k}},\bm{y^{k^{'}}} \right )}{\sqrt{dVar\left ( \bm{y^{k}} \right )dVar\left ( \bm{y^{k^{'}}} \right )}},
\end{equation}
where $\bm{y^k}$ and $\bm{y^{k^{'}}}$ are respectively the $k$ and $k^{'}$-th chunks of the vector $y$; $dCov(\cdot)$ represents the distance covariance between two matrices; $dVar(\cdot)$ is the distance variance of each matrix. In this way, we can define the regularization losses $L_p, L_q, L_t, L_v$ to encourage the feature independence of the factors in user ID, item ID, textual feature and visual feature, respectively. 

Finally, the regularization loss for disentangled learning is formulated as:
\begin{equation}
	L_d = L_p + L_q + L_v + L_t.
\end{equation}
Note that in this equation, the weights of different regularization losses ($L_p, L_q, L_v, L_t$) can be empirically tuned. Here, we set them to be equal in the learning process for simplicity.

\subsubsection{\textbf{Modality Preference Modeling}}
\label{UDP}
Typically, for a certain factor, users often value its features exhibited by different modalities differently. For example, for the \emph{appearance} of a product, a user will take the visual features (from images) more seriously than the textual features (from reviews); and  value the textual features more than the visual features for the \emph{quality} of the product. To capture the complicated user modality preferences on various modalities for each factor, we design a weight-shared multimodal attention mechanism in DMRL. Specifically, for an item $i$, a user $u$'s specific attention $a_{u,i}^k$ to different modalities for the $k$-th factor is computed as:
\begin{equation}
	\label{eq:attention}
	\bm{h_{u,i}^k} = Tanh( \bm{W} [ \bm{p}_u^k;\bm{q}_i^k;\bm{q}_t^k;\bm{q}_v^k] + \bm{b} ),
\end{equation}
\begin{equation}
	\bm{\hat{a}_{u,i}^k} = \mathbf{W}_{v} \bm{h_{u,i}^k},
\end{equation}
where $\bm{W}$ and $\bm{b}$ are respectively the weight matrix and bias vector of the neural network. In order to enhance efficiency of learning, we use shared weight under all factors. $\bm{h_{u,i}^k}$ is an output of of the hidden layer; $\mathbf{W}_{v}$ is a transformation matrix that projects the hidden layer into an output attention weight vector. $[  \bm{p}_u^k;\bm{q}_i^k;\bm{q}_t^k;\bm{q}_v^k]$ denotes the concatenation of $\bm{p}_u^k$, $\bm{q}_i^k$, $\bm{q}_t^k$ and $\bm{q}_v^k$. \emph{Tanh} is used as the activation function.

Following the standard procedure of attention mechanism, there is a subsequent step to normalize $\bm{\hat{a}_{u,i}^k}$ with the softmax function, which converts the attention weights to a probability distribution. In our model, the final attention weight for the textual feature of the $k$-th factor is computed as:
\begin{equation}
	a_{u,i,t}^k = \frac{exp\left ( \hat{a}_{u,i,t}^k \right )}{exp\left ( \hat{a}_{u,i,q}^k \right )+exp\left ( \hat{a}_{u,i,t}^k \right )+exp\left ( \hat{a}_{u,i,v}^k \right )}.
\end{equation}
Similarly, the weight for item $i$'s ID and visual feature $a_{u,i,q}^k$ and $a_{u,i,v}^k$ can be computed in the same way.

\subsubsection{\textbf{Preference Prediction}}
User's preference for an item should be estimated based on her preferences on different modalities of all factors for this item. Hence, with the representations of different modalities for different factors, given a user $u$ and a target item $i$ with their representations, the user preference on the $k$-th factor according to the features of the $x$-modality is estimated as:
\begin{equation}
	\label{upm:dot_product}
	r_{u,i,x}^k =  a_{u,i,x}^k \cdot \sigma(\bm{{p_u^k}}\odot\bm{q_x^k}).
\end{equation}
where $\odot$ denotes dot product. Notice that the attention weight $a_{u,i,x}^k$ represents the importance of the $x$-modality to the $k$-th factor for the user's preference to the item; and the dot product between $\bm{{p_u^k}}$ and $\bm{q_x^k}$ estimates how the features of the $x$-modality fits this user's tastes on the $k$-th factor.  The integration of both terms can evaluate the user's preference to the item on the $k$-th factor according to the features of $x$-modality more comprehensively. $\sigma(\cdot)$ is the activation function and softplus is used to ensure that the resultant score is positive.
Similarly, we can compute $r_{u,i,q}^k$, $r_{u,i,t}^k$ and $r_{u,i,v}^k$ in the same way, which represent the user preferences on the $k$-th factor from different modalities, i.e., item ID, reviews, and images, respectively. 
And the final score of user preference on the k-th factor of the target item is predicted  by aggregating the scores from different modalities:
\begin{equation}
	\label{upm:aggregation}
	r_{u,i}^k =  r_{u,i,q}^k + r_{u,i,t}^k + r_{u,i,v}^k.
\end{equation}

Finally, the predicted scores of all the factors are integrated together to predict  user preference to the target item, which is:
\begin{equation}
	r_{u,i} =  \sum_{k=1}^{K}r_{u,i}^k.
\end{equation}

\subsection{Model Learning}
\subsubsection{Objective function}
We target at the top-$n$ recommendation, namely, to recommend a set of $n$ top-ranked items which match the target user's preferences. Similar to other rank-oriented recommendation studies~\cite{zhang2017joint,wang2019ngcf}, we use the pairwise-based learning method for optimization. The loss function is defined as:
\begin{equation} \label{eq:obj}
	L_{BPR} = \sum_{(\mathbf{u}, \mathbf{i}^+,\mathbf{i}^-)\in{\mathcal{O}}} -\ln\phi(r_{u,i} - r_{u,i^-}),
\end{equation}
where $\mathcal{O}=\{(u, i^+, i^-)|(u,i^+)\in\mathcal{R^+}, (u,i^-) \in\mathcal{R^-}\}$ denotes the training set;  $\mathcal{R^+}$ indicates the observed interactions between user $u$ and $i^+$ in the training dataset, and $\mathcal{R^-}$ is the sampled unobserved interaction set. For each positive pair $(u, i^{+})$, we first randomly sample $n$ negative ones from the items that a user hasn't purchased before as candidate items~\footnote{In our implementation, we empirically set $n=4$ to balance efficiency and effectiveness.}. From the candidate items, we select the one which is most similar to the user $u$ based on the dot product of their embeddings, as a hard negative sample to construct the negative pair $(u, i^{-})$.


\subsubsection{Optimization}
With the consideration of all the regularization terms, the final object function of our DMRL is,
\begin{equation}
	\begin{split}
		loss = L_{BPR} + \lambda _{\theta}L_{\theta} + \lambda _{d}L_{d},
	\end{split}
\end{equation}
where $L_{\theta} = \left\|\Theta\right\|^2_2$ represents the $L_2$ regularization for the parameters $\Theta$ of the model; $\lambda_{\theta}$ and $\lambda_{d}$ are hyperparameters that control the weights of $L_2$ regularizer and independence regularizer, respectively. The optimization is quite standard and the stochastic gradient descent (SGD) algorithm is adopted. In implementation, the Adam optimizer~\cite{kingma2014adam} is adopted to tune the learning rate.

\section{Experiments}
\label{sec:experiments}






To validate the effectiveness of our model, we conducted extensive experiments on five public datasets. In the next, we first introduce the experimental setup, and then report and analyze the experimental results.

\vspace{0pt}
\subsection{Experimental Setup}
\begin{table}[t]
	\centering
	\caption{ Basic statistics of the used datasets.}
	\label{tab:data}
	\begin{tabular}{ccccl}
		\toprule
		Dataset&\#user&\#item&\#interactions&sparsity \\
		\midrule
		Office Products& 4,874& 7,279 & 52,957 &99.85\% \\
		Baby & 12,637& 18,646 & 121,651 &99.95\% \\
		Clothing & 18,209& 35,526 & 150,889 &99.98\% \\
		Toys Games & 18,748& 30,420 & 161,653 &99.97\% \\ 
		Sports & 21,400& 36,224 & 982,618 &99.97\% \\
		\bottomrule
	\end{tabular}
	\vspace{0pt}
\end{table}
\subsubsection{\textbf{Datasets}}
The public Amazon review dataset~\footnote{http://jmcauley.ucsd.edu/data/amazon.}~\cite{mcauley2013hidden}, which has been widely used for recommendation evaluation in previous studies, is used for evaluation in our experiments. This dataset contains user interactions (review, rating, helpfulness votes, etc.) on items as well as the item metadata (descriptions, price, brand, image features, etc.) on 24 product categories. Five product categories in this dataset are selected and all the reviews and item images are kept as side information for items. We pre-processed the dataset to keep only the items
and users with at least 5 interactions. The basic statistics of the five datasets are shown in Table~\ref{tab:data}. For each observed user-item interaction, we treated it as a positive instance, and then paired it with negative items which are randomly sampled from items that the user has not purchased before. 

In this work, we focus on the top-$n$ recommendation task, which aims to recommend a set of top-$n$ ranked items that will be appealing to the target user. For each dataset, we randomly selected 80\% of the interactions from each user to construct the training set, and the remaining 20\% for testing. From the training set, we randomly
selected 10\% of interactions as a validation set to tune hyper-parameters. 
\begin{table*}[ht]
	\caption{Performance of our DMRL model and the competitors over five datasets. The best results are highlighted in bold.} 
	\centering
	\begin{tabular}{l|cc|cc|cc|cc|cc}
		\hline
		Datasets&\multicolumn{2}{c|}{Office Products}&\multicolumn{2}{c|}{Baby}&\multicolumn{2}{c|}{Clothing}&\multicolumn{2}{c|}{Toys Games}&\multicolumn{2}{c}{Sports}\\
		\hline
		Metrics & Recall & NDCG  
		& Recall & NDCG 
		& Recall & NDCG  
		& Recall & NDCG
		& Recall & NDCG\\
		\hline\hline
		NeuMF &0.0605	&0.0337	&0.0502	&0.0224	&0.0189	&0.0080	&0.0253	&0.0128	&0.0330	&0.0157\\
		CML &0.1229	&0.1395	&0.0674	&0.0444	&0.0409	&0.0224	&0.1227	&0.1160	&0.0982	&0.0640\\
		NGCF &0.0814	&0.0479	&0.0694	&0.0313	&0.0466	&0.0216	&0.0970	&0.0587	&0.0707	&0.0337\\
		DGCF &0.1206	&0.1105	&0.0788	&0.0465	&0.0745	&0.0405	&0.1262	&0.1085	&0.1026	&0.0629\\
		\hline
		JRL &0.0656	&0.0365	&0.0579	&0.0266	&0.0233	&0.0124	&0.0472	&0.0413	&0.0368	&0.0214\\
		MMGCN & 0.1173	& 0.1231	& 0.0814	& 0.0496	& 0.0573	& 0.0312	& 0.1171	 & 0.1056	& 0.0913	& 0.0572 \\
		MAML  & 0.1333	& 0.1412	& 0.0867	& 0.0521	& 0.0782	& 0.0387	& 0.1183	 & 0.1117	& 0.1029	& 0.0676 \\
		GRCN &\underline{0.1351}	&\underline{0.1524}	&\underline{0.0883}	&\underline{0.0541}	&\underline{0.0861}	&\underline{0.0452}	&\underline{0.1336}	& \underline{0.1236} & \underline{0.1065} & \underline{0.0693} \\ 
		DMRL &\textbf{0.1563}&\textbf{0.1842}&\textbf{0.0906}&\textbf{0.0561}&\textbf{0.0917}&\textbf{0.0511}&\textbf{0.1434}&\textbf{0.1331}&\textbf{0.1111}&\textbf{0.0711}\\
		\hline
	\end{tabular}
	\label{tab:results}
	\vspace{0.0cm}
\end{table*}
\subsubsection{\textbf{Baselines and Evaluation Metrics}}
We compare our DMRL model with the following baselines, including both deep (NeuMF~\cite{he2017neural}, JRL~\cite{zhang2017joint}) MF and metric learning (CML~\cite{hsieh2017cml}, MAML~\cite{liu2018MAML}) based models, as well as the graph based methods (NGCF~\cite{wang2019ngcf}, DGCF~\cite{wang2020DGCF}, MMGCN~\cite{wei2019mm}, GRCN~\cite{Wei2019GRCN}). 

For each user in the test set, we treat all the items that the user did not interact with as negative items. Two widely used metrics for top-$n$ recommendation are adopted in our evaluation: \emph{Recall} and \emph{Normalized Discounted Cumulative Gain} (NDCG)~\cite{he2015trirank}.
For each metric, the performance is computed based on the top 20 results. Notice that the reported results are the average values across all the testing users.

In experiments, for all the baselines, we use the codes they released for evaluation. And, we put great efforts to tune hyperparameters of these methods and reported their best performance.

\subsubsection{\textbf{Parameter Settings}}
We implemented our model with Tensorflow~\footnote{https://www.tensorflow.org.} and carefully tuned the key parameters. The embedding size is fixed to 128 for all methods and the embedding parameters are initialized with the Xavier method~\cite{Xavier2010xavier}. We optimized our method with Adam~\cite{kingma2014adam} and used the default learning rate of 0.0001 and default mini-batch size of 1024. The $L_{2}$ regularization coefficient $\lambda_{\theta}$ and the distance correlation $\lambda_{d}$ are searched in the range of $\{1e^{-5},1e^{-4},\cdots,1e^{+1}\}$. We search the number of the factors being disentangled in $\{1,2,4,8\}$. Besides, model parameters are saved in every 10 epochs. The early stopping strategy~\cite{wang2019ngcf} is performed, $i.e.$, premature stopping if recall@20 does not increase for 50 successive epochs. In addition, our codes are released to facilitate the replication of our experiments~\footnote{https://github.com/liufancs/DMRL.}.

\subsection{\textbf{Performance comparison}}

The results of our model and all the competitors over the five datasets are reported in Table~\ref{tab:results}. Overall, it can be seen that our method outperforms all the competitors consistently across all the datasets in terms of different metrics. By grouping all the methods into two categories based on whether they use multimodal information, we make some interesting observations. 

The methods in the first block only use the user-item interactions. NeuMF models the non-linear interactions between users and items by using deep neural networks and achieves better performance than the traditional MF methods using linear interactions~\cite{he2017neural}. NGCF exploits high-order connectivities between users and items through embedding propagation over the graph structure. Hence it obtains better user and item representations than NeuMF. DGCF outperforms NGCF across all the datasets. This is attributed to the utilization of disentangled representation to learn robust and independent user and item embeddings by considering users' diverse intents. However, the aforementioned methods all use the dot product as the interaction function. As pointed out in~\cite{hsieh2017cml}, dot product does not obey the triangle inequality and thus cannot model fine-grained user preferences well. 
By using a metric-based learning approach, CML outperforms both NeuMF and NGCF by a large margin. This observation is also consistent with the results reported in~\cite{liu2018MAML}. 

All the methods in the second block exploit both textual and visual features besides the interaction information. In general, these methods yield better performance than those without using multimodal features, demonstrating the effectiveness of leveraging side information on modeling user preference. Owing to the valuable information in multimodal features, JRL outperforms NeuMF accross all the datasets. By exploiting user-item interactions to guide the representation learning in different modalities, MMGCN yields better performance over NGCF on all the datasets. However, MMGCN underperforms DGCF on four datasets, which is because the entangled multimodal representations limit the capability of representation learning. MAML outperforms MMGCN across all the datasets, owing to the adoption of metric-based learning approach and attention mechanism to capture user diverse preferences. By discovering and pruning potential false-positive edges, GRCN obtains better performance across all datasets.  

DMRL outperforms all the baselines consistently over all the datasets. We credit this to the joint effects of the following four aspects. Firstly, the utilization of multimodal information on modeling user preference, which can be observed from the performance comparisons of methods between two blocks in Table~\ref{tab:abresults}. Secondly, DMRL models user preference by considering the differing contributions of different modalities to each factor. Thirdly, an attention network is designed in our model to capture user's modality preference on different modalities for each factor. Finally, the use of the disentangled representation learning technique learns independent representations for different factors. As demonstrated in DGCF, disentangled representation can better model users' multiple intents and thus achieve better performance.


\begin{table*}[t]
	\caption{Performance of our DMRL model and its variants over five datasets. The best results are highlighted in bold. } 
	\centering
	\begin{tabular}{l|cc|cc|cc|cc|cc}
		\hline
		Datasets&\multicolumn{2}{c|}{Office}&\multicolumn{2}{c|}{Baby}&\multicolumn{2}{c|}{Clothing}&\multicolumn{2}{c|}{ToysGames}&\multicolumn{2}{c}{Sports}\\
		\hline
		Metrics & Recall & NDCG  
		& Recall & NDCG 
		& Recall & NDCG  
		& Recall & NDCG
		& Recall & NDCG\\
		\hline\hline
		DGCF &0.1206 &0.1105	&0.0788	&0.0465	&0.0745	&0.0405	&0.1262	&0.1085	&0.1026	&0.0629\\ 
		DMRL$_{t}$ &0.1517	&0.1578	&0.0783	&0.0497	&0.0766	&0.0419	&0.1258	&\underline{0.1257}	&0.1071	&\underline{0.0657}\\
		DMRL$_{v}$ &0.1532	&\underline{0.1612}	&\underline{0.0852}	&\underline{0.0534}	&\underline{0.0897}	&\underline{0.0464}	&\underline{0.1395}	&0.1287	&\textbf{0.1122}	&0.0682\\
		\hline
		DMRL$_{w/o\ a}$ &0.1513 &0.1454 &0.0759 &0.0479 &0.0781 &0.0392 &0.1363 &0.1262 &0.0998 &0.0597 \\
		DMRL$_{w/o\ u}$ &\underline{0.1541}	&0.1559	&0.0835	&0.0513	&0.0841	&0.0427	&0.1211	&0.1186	&\underline{0.1112}	&0.0691\\
		DMRL &\textbf{0.1563}&\textbf{0.1842}&\textbf{0.0906}&\textbf{0.0561}&\textbf{0.0917}&\textbf{0.0511}&\textbf{0.1434}&\textbf{0.1331}&0.1111& \textbf{0.0711}\\
		\hline
	\end{tabular}
	\label{tab:abresults}
	\vspace{0.0cm}
\end{table*}

\subsection{\textbf{Effects of Modality Preference}}
One of our assumptions in the model is that for each factor of an item, different users may value its features of one modality more than those of another modality. To capture this, we design an attention mechanism to compute the attention weights for each factor of different modalities. The other assumption is that different modalities contribute differently for a user's preference  to an item. With this consideration, for each user-item pair, our model computes the rating given to each factor of an item across different modalities (see Eq.~\ref{upm:dot_product}). In this section, we would like to validate the above two assumptions via visualization: 1) visualizing the learned attention weights to illustrate user's modality preferences on different modalities for each factor; and 2) visualizing the user preferences (ratings) to illustrate two users' preferences on different factors represented by different modalities.
\begin{figure}[t]
	\centering
	\subfloat[User A]{\includegraphics[width=0.5\linewidth]{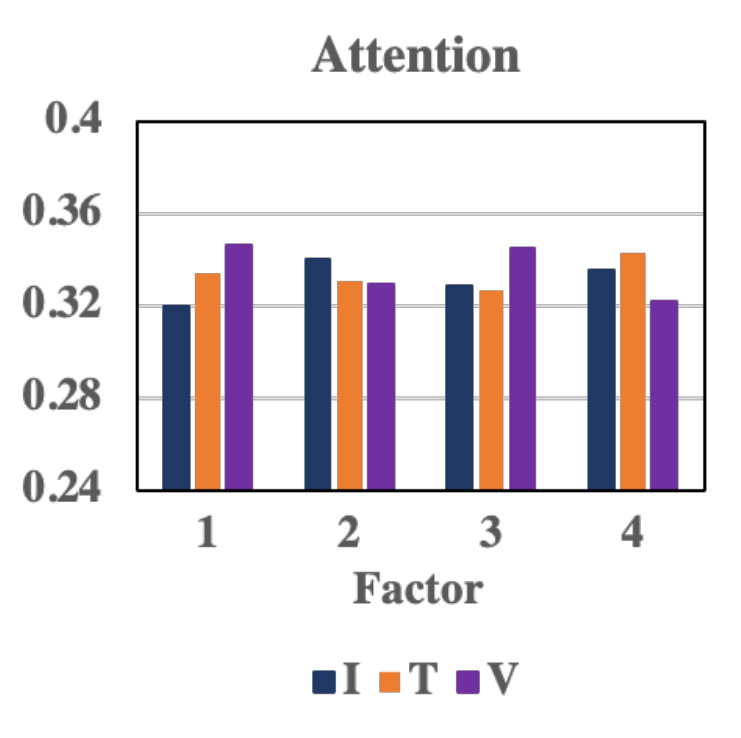}}
	\subfloat[User B]{\includegraphics[width=0.5\linewidth]{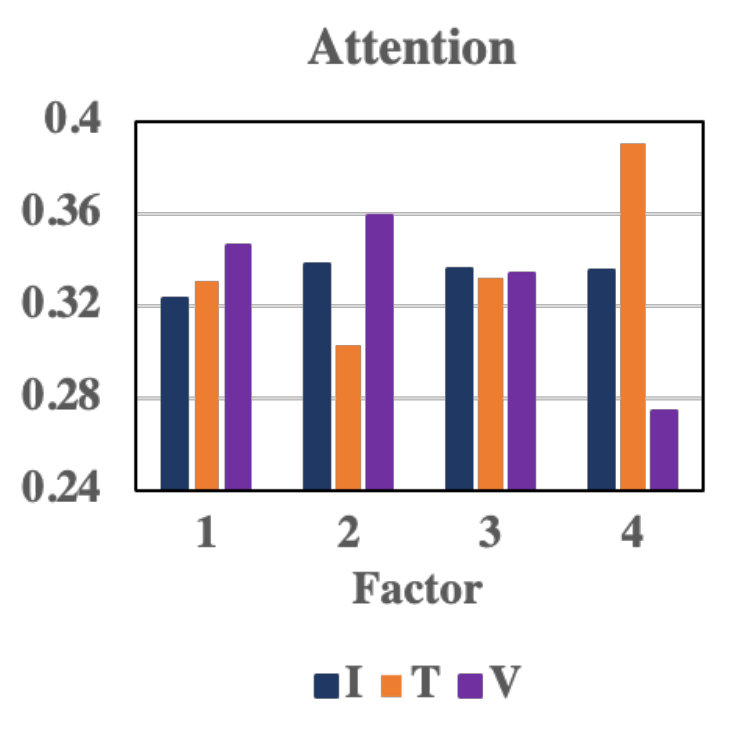}}
	\caption{Attention weights of different modalities. I, T and V represent item ID, textual feature and visual feature, respectively.}
	\label{fig:Attention12}
\end{figure}
\begin{figure}[t]
	\centering
	\subfloat[User A]{\includegraphics[width=0.5\linewidth]{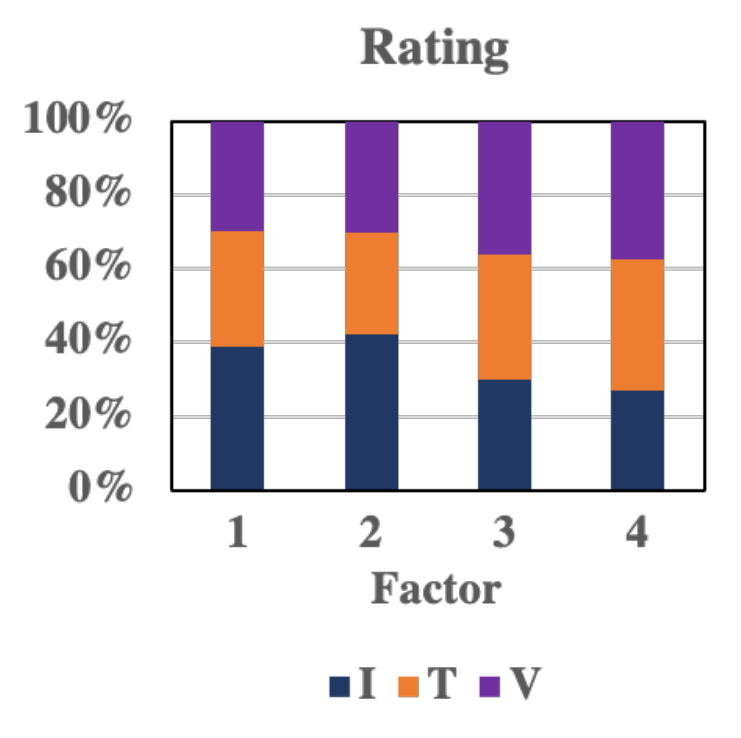}}
	\subfloat[User B]{\includegraphics[width=0.5\linewidth]{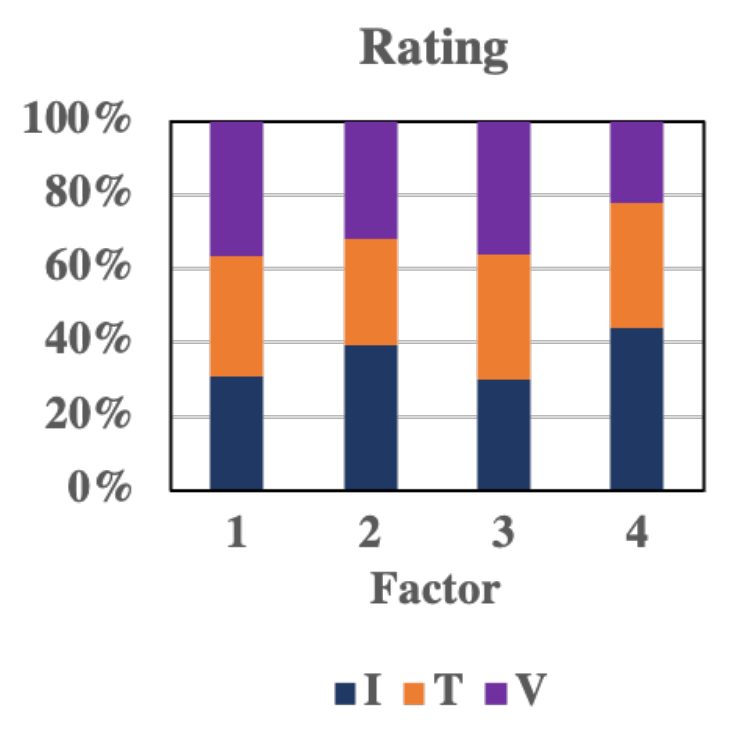}}
	\caption{Ratings from two users. I, T and V represent item ID, textual feature and visual feature, respectively.}
	\label{fig:Rating12}
\end{figure}
\begin{figure}[t]
	\centering
	\subfloat[Entangled Vector]{\includegraphics[width=0.5\linewidth]{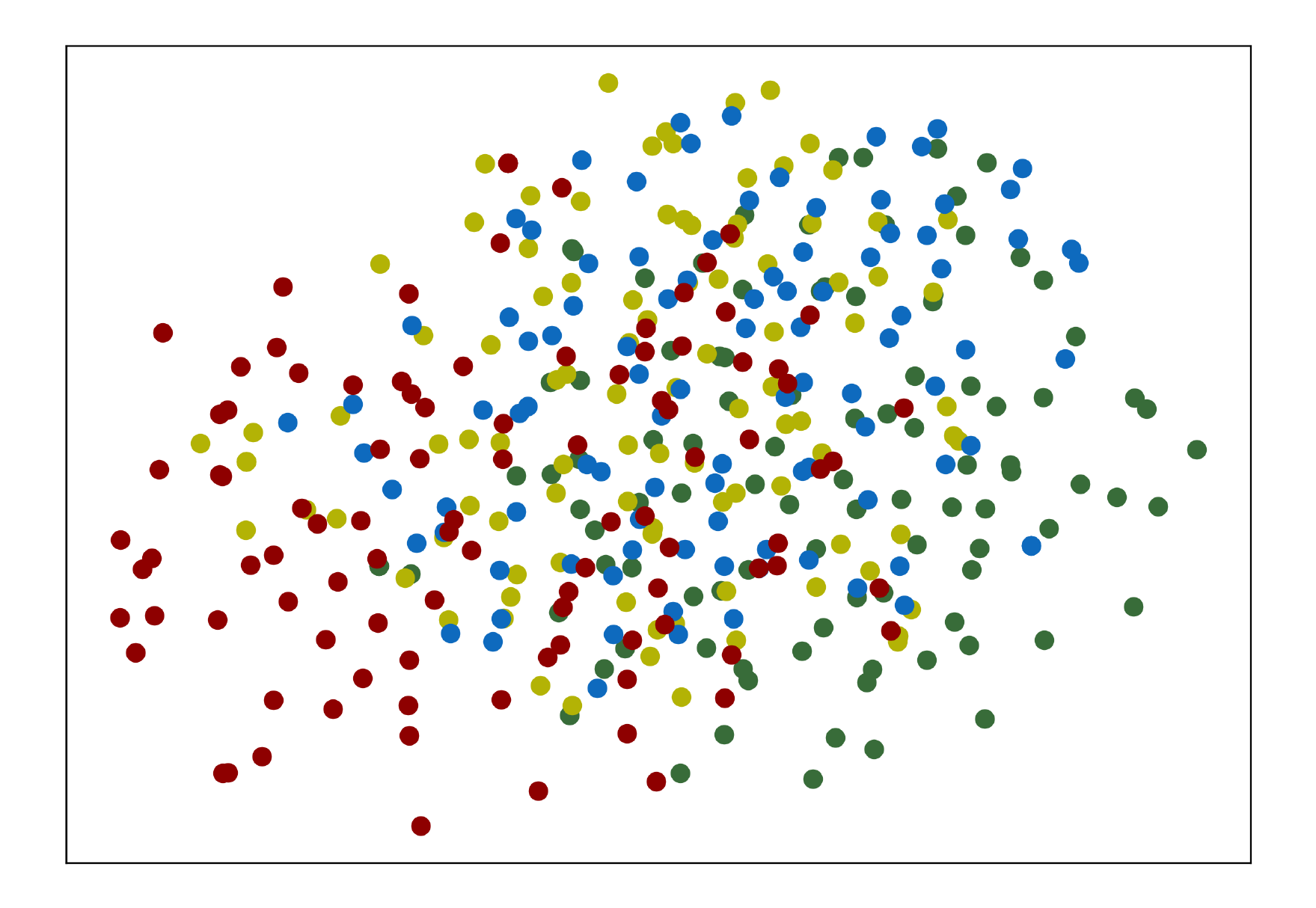}}
	\subfloat[Disentangled Vector]{\includegraphics[width=0.5\linewidth]{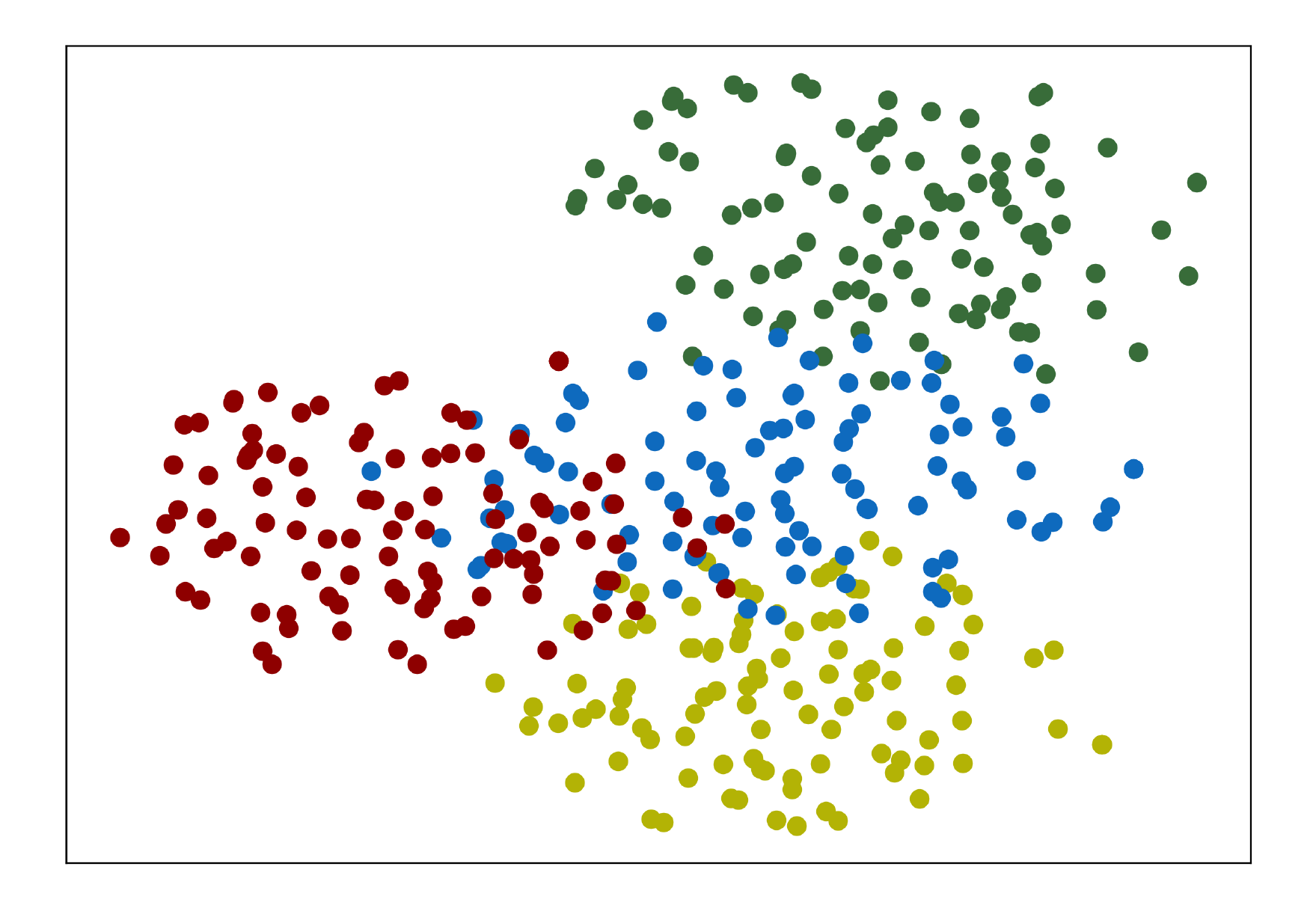}}
	\caption{Visualization of entangled and disentangled vectors extracted from textual features. Best viewed in color.}
	\label{fig:visualization}
\end{figure}
\subsubsection{\textbf{Modality Preferences (Attention Weights)}}
\label{E:modalitypreference}
In order to study the user's modality preferences on different modalities for each factor, we  compute the attention weights for different modalities of each factor (in the \emph{Office} dataset)~\footnote{We selected two users (0 and 197) who purchased the same item (1294) as an example.},which are visualized in Fig.~\ref{fig:Attention12}. Specifically, I, T and V represent item ID, textual feature and visual feature, respectively. The $x$-axis represents different factors. Comparing the results in Fig.~\ref{fig:Attention12}(a) and Fig.~\ref{fig:Attention12}(b), we can see that the weights of different modalities for the same factor are different between the two users. 
This verifies that for an item, the preferences of users to different modalities are different for each factor. In addition, it should be noted that the attention weight cannot represent the contribution of a modality to a user's preference to a factor of the target item. 

\subsubsection{\textbf{User Preferences (Ratings)}}
\begin{figure*}[t]
	\centering
	\subfloat[Recall on Office]{\includegraphics[width=0.2\linewidth]{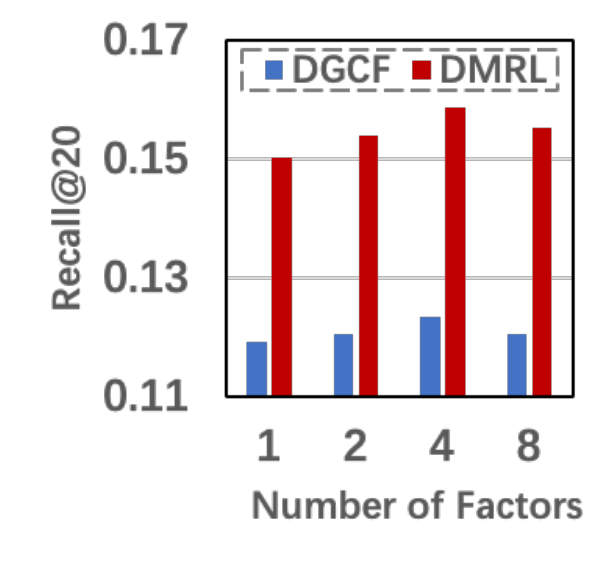}}
	\subfloat[Recall on Baby]{\includegraphics[width=0.2\linewidth]{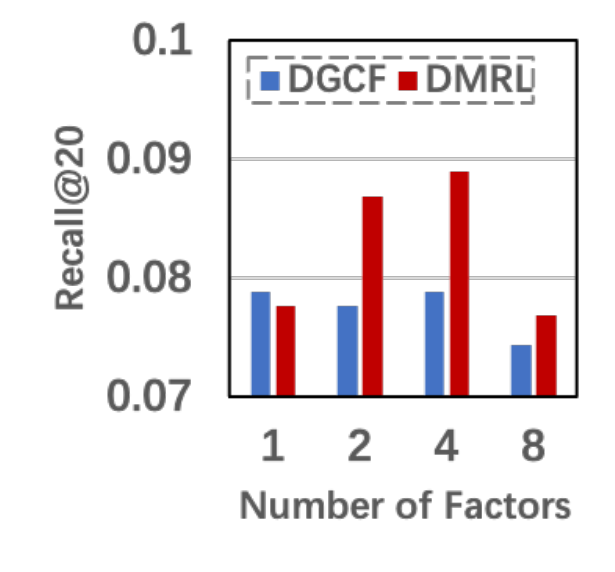}}
	\subfloat[Recall on Clothing]{\includegraphics[width=0.2\linewidth]{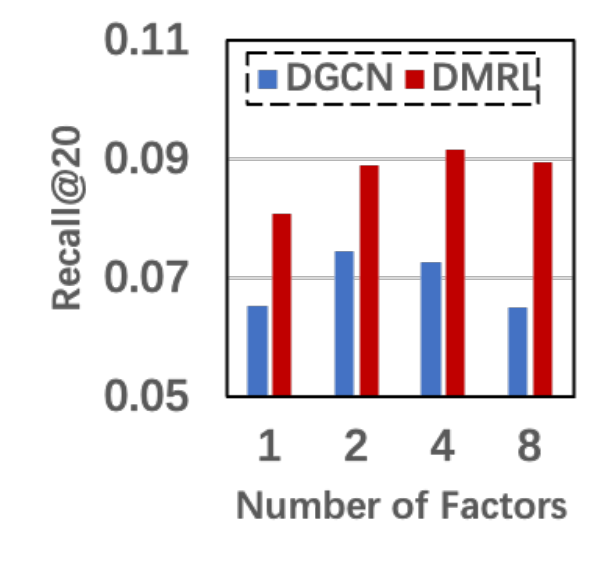}}
	\subfloat[Recall on ToysGames]{\includegraphics[width=0.2\linewidth]{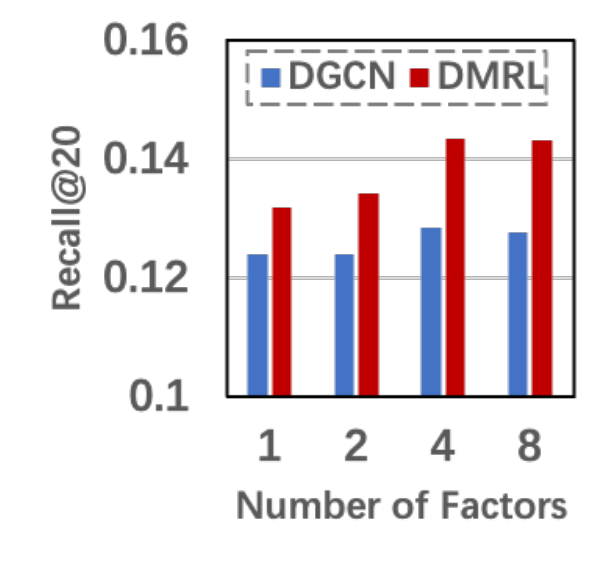}}
	\subfloat[Recall on Sports]{\includegraphics[width=0.2\linewidth]{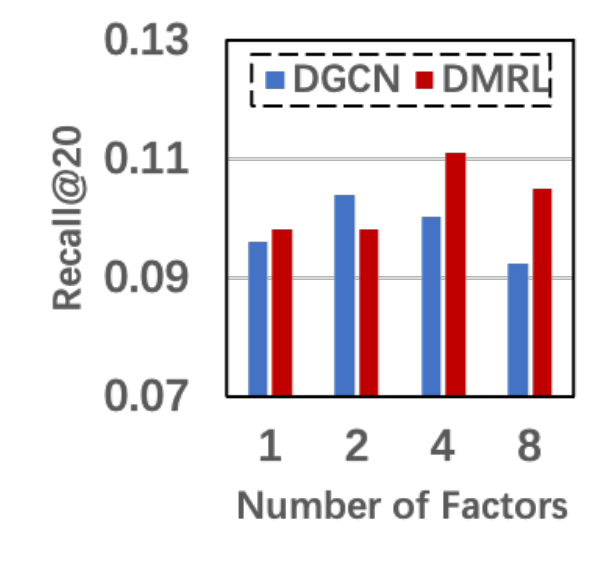}}\\
	\subfloat[NDCG on Office]{\includegraphics[width=0.2\linewidth]{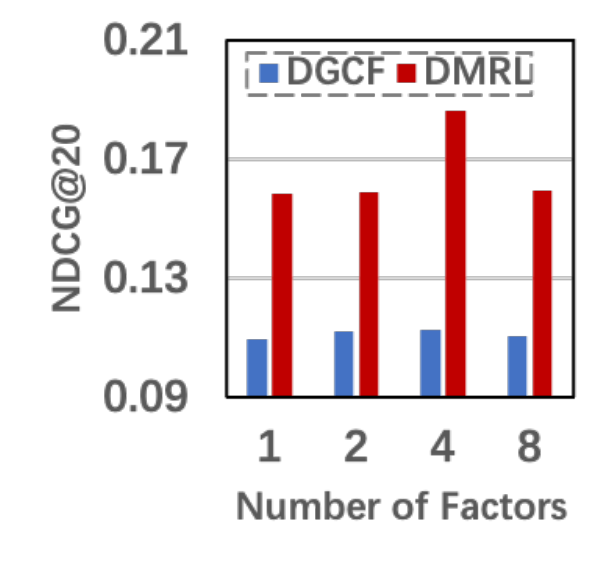}}
	\subfloat[NDCG on Baby]{\includegraphics[width=0.2\linewidth]{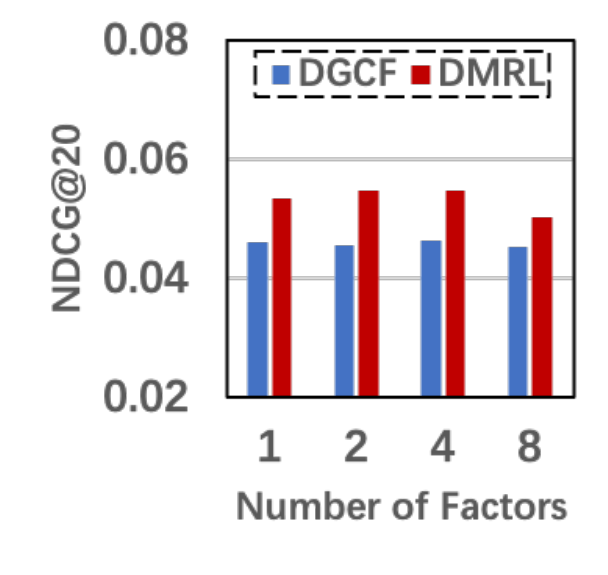}}
	\subfloat[NDCG on Clothing]{\includegraphics[width=0.2\linewidth]{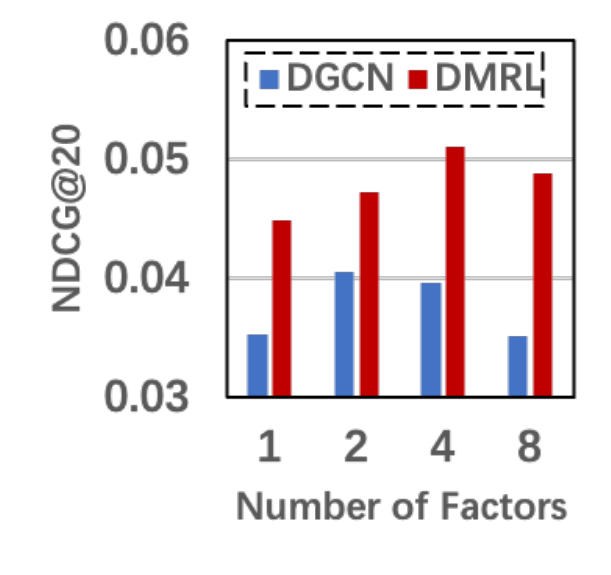}}
	\subfloat[NDCG on ToysGames]{\includegraphics[width=0.2\linewidth]{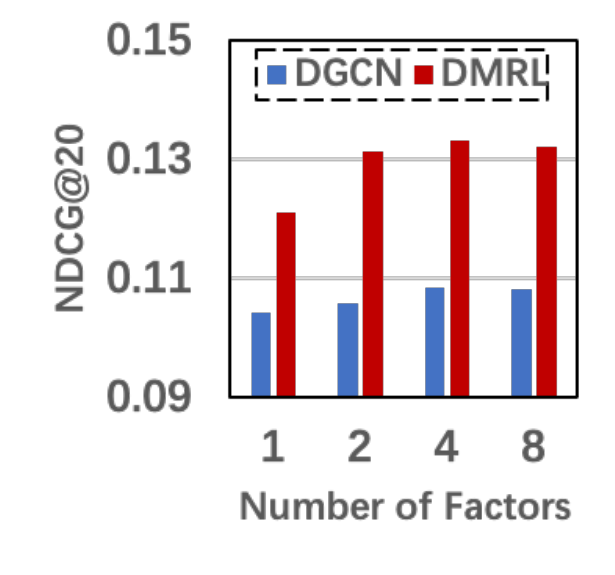}}
	\subfloat[NDCG on Sports]{\includegraphics[width=0.2\linewidth]{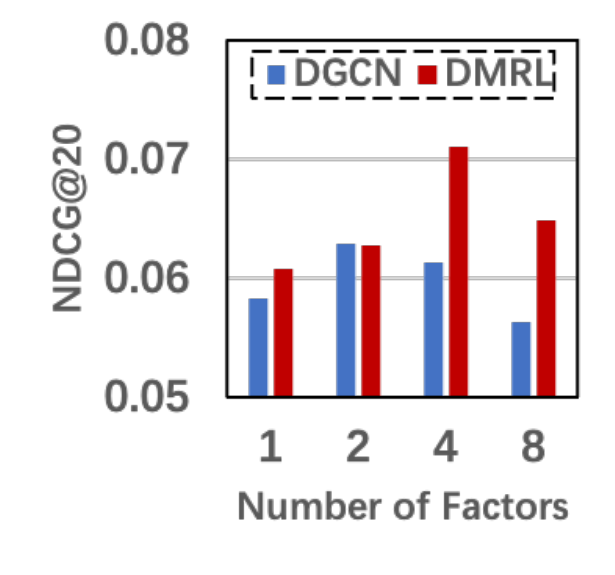}}
	\caption{Impact of the Factor Number (\textit{K}).}
	\label{fig:disentangle}
\end{figure*}

In order to further study the contributions of different modalities to a user's preference on the factors of the target item, we compute the ratings of the two different users given to different factors of the same item represented by different modalities. The ratings are then normalized across different modalities and visualized in Fig.~\ref{fig:Rating12}.  Overall, both Fig.~\ref{fig:Rating12}(a) and Fig.~\ref{fig:Rating12}(b) show that for the same factor, the features of different modalities have different contributions to user preference. For example, item ID yields a higher score for the second factor than the scores of both visual feature and textual feature. In other words, the textual and visual features contribute relatively less to the user's preference on this factor. What's more, take the results in the 4-$th$ factor in Fig.~\ref{fig:Rating12}(b) and Fig.~\ref{fig:Attention12}(b) for example, we can find that although the user pays more attention to the textual feature according to the attention weight, the item ID actually contributes more to the user preference. 

\subsection{\textbf{Effect of Disentanglement}}
In Fig.~\ref{fig:visualization}, we visualized the entangled and disentangled vectors extracted from textual features. The textual features is associated with the items from \emph{Office}~\footnote{We randomly sampled the items and their associated textual features.}. In both Fig.~\ref{fig:visualization}a and Fig.~\ref{fig:visualization}b, the dots with the same color (e.g., red) denote the vectors of the same factor. We use t-SNE~\cite{vanDerMaaten2008} for clustering and visualization. The clustering results based on the vectors of different factors (right figure) can illustrate the effectiveness of using disentangled representation technique. In other words, it demonstrates that our proposed method could encourage independence of the vectors of different factors for modality features.

\subsection{\textbf{Ablation Study}}
\label{sec:ablation}

In this section, we examine the contribution of different components to the performance of our model. Our analyses are based on the performance comparisons to the following methods.
\label{sec:visualization}
\begin{itemize}[leftmargin=*]
	\item \textbf{DGCF~\cite{wang2020DGCF}}: It is a disentangled representation based method which only uses item IDs. 
	\item \textbf{DMRL$_{t}$}: It is a variant of our method which only uses item IDs and textual features.
	\item \textbf{DMRL$_{v}$}: It is a variant of our method which only uses item IDs and visual features.
	\item \textbf{DMRL$_{w/o\ a}$}: This variant removes the designed multimodal attention mechanism. It exploits the multimodal features of different factors indiscriminately.
	\item \textbf{DMRL$_{w/o\ u}$}: This variant removes user information in Eq.~\ref{eq:attention}, namely, for the same item, the attention weights are all the same for different users. It is designed to investigate the effectiveness of modeling user modality preference on different modalities by comparing to our DMRL model.
\end{itemize}

The results of those variants and our method are reported in Table~\ref{tab:abresults}. We group all the methods and variants into two groups based on whether the methods use all the modalities. The methods in the first block only use a part of three modalities. Both DMRL$_t$ and DMRL$_v$ outperform DGCF, demonstrating the effectiveness of leveraging side information to learn user preference. DMRL$_v$ outperforms DMRL$_t$ in terms of Recall@20 over all the datasets, which indicates the importance of visual features on user preference modeling. DMRL$_t$ performs better than DMRL$_v$ in terms of NDCG@20 on \emph{ToysGames} and \emph{Sports}. This is because textual features have a higher contribution for the two cases. 

The variants in the second block use features of all three modalities. DMRL$_{w/o\ u}$ outperforms DMRL$_{w/o\ a}$ in almost all the cases, which indicates the importance of distinguishing different contributions of different modalities. More importantly, without the attention mechanism, DMRL$_{w/o\ a}$ even underperforms DMRL$_v$ and DMRL$_t$, which use our attention method. This further demonstrates the importance of modeling users' modality preferences. The further improvement of DMRL over DMRL$_{w/o\ u}$ validate the effectiveness of distinguishing different users' attentions to various modalities for each factor. 

It is surprising that the methods in the second block underperform the methods in the first block on $Sports$ in terms of Recall@20. This warns us that it is possible that the features of different modalities are inconsistent, which may exert negative influence on preference modeling. More advanced multimodal feature fusion methods are needed to tackle this problem. Besides this special case, DMRL can achieve superior performance over DMRL$_v$ and DMRL$_t$ consistently, which indicates the benefits of considering multimodal information in recommendation. The comprehensive ablation studies also demonstrate the positive utility of different components of our DMRL model.

\subsection{\textbf{Effects of the Factor Number}}
\label{sec:missingatt}

In this section, we study the influence of the factor number on the performance of our method. 
To compare the performance of our model and the baseline model with different factor numbers, we carry out experiments by tuning the number of factors in $\{1, 2, 4, 8\}$ on different datasets.  

We compare our model to DGCF which also uses disentangled representations but did not use multimodal features. Fig.~\ref{fig:disentangle} shows the results in terms of Recall@20 and NDCG@20 on all five datasets. As we can see, with the increasing of the number of factors, both DMRL and DGCF achieve better performance and the best performance is obtained when the number is 2 or 4. This indicates that recommendation methods can benefit from robust and independent representation learning technique. Specifically, DGCF performs worse over \emph{Office}, \emph{Baby} and \emph{Clothing} when $\textit{K} = 1$ and $\textit{K} = 2$, indicating that insufficient number of factors limits the capability of disentangled representation. However, the recommendation performance drops when when $\textit{K} = 8$. The reason might be that the chunked representations with small embedding size have limited expressiveness in representation learning for each factor~\cite{wang2020DGCF}. The large improvement over DGCF shows the effectiveness of our model on exploiting multimodal features.

\section{Conclusion}
\label{sec:conclusion}
In this work, we present a novel recommendation model called Disentangled Multimodal Representation Learning (DMRL), which models user's modality preference with respect to multimodal information on different factors of items. In DMRL, we employ disentangled representation learning technique to disentangle the representations of different factors in each modality. In addition, we design a multimodal attention mechanism to capture users' modality preferences to different modalities for each factor, so as to learn better representations for recommendation. Finally, we estimate a user's preference score to each factor of the target item based on its representation in each modality, and then combine the scores together across all modalities and factors with the computed attention weights. Extensive experiments on five publicly available datasets show that our model outperforms the state-of-the-art methods, demonstrating the superiority of our method in exploiting multimodal information. The ablation studies further validate the importance of modeling personal modality preference towards different modalities.

\ifCLASSOPTIONcompsoc
\else
\fi

\ifCLASSOPTIONcaptionsoff
\fi

\bibliographystyle{IEEEtran}
\bibliography{DMRL}

\end{document}